\documentclass[aps,prd,twocolumn,superscriptaddress,nofootinbib,longbibliography,floatfix]{revtex4-2}
\usepackage{bbm}
\usepackage{amsmath,amssymb,bm,graphicx,hyperref}
\usepackage[dvipsnames]{xcolor}
\usepackage{amsthm}
\usepackage{float}
\usepackage{hyperref}
\usepackage{dsfont}
\usepackage{xspace}
\newcommand{\one}{\mathds{1}}

\newtheorem{theorem}{Theorem}
\newcommand{\g}{\gamma}
\newcommand{\gfive}{\g_5}
\newcommand{\Deff}{D_{\mathrm{eff}}}
\newcommand{\Tr}{\operatorname{Tr}}
\newcommand{\anticomm}[2]{\{#1,#2\}}
\newcommand{\CDF}{CDF\xspace}
\newcommand{\IMF}{IMF\xspace}

\begin{document}

\title{Exact chiral symmetry, non-Hermiticity, and the axial anomaly on the lattice:\\ A symmetry-class criterion for disorder-induced undoubling}

\author{Laurence A.\ Jacobs}

\affiliation{University of Zurich, Zurich, Switzerland}
\affiliation{Center for Complexity Sciences, National University of Mexico, Mexico City, Mexico}

\date{November 10, 2026 \\[0.5em] \mbox{\url{https://doi.org/10.5281/zenodo.XXXXX}}}

\begin{abstract}
A lattice Dirac operator built to keep chiral symmetry exact by weighting the derivative with a quenched random field, the \emph{chiral disordered fermions} (\CDF) of [L.\ Jacobs, Phys.\ Rev.\ Lett.\ \textbf{51}, 172 (1983)], is reexamined with tools unavailable when it was first formulated. We show that the disorder-averaged operator satisfies $\anticomm{\Deff}{\gfive}=0$ \emph{exactly, to all orders} in the disorder strength: the disorder lifts the doubler degeneracy, pushing the doublers off the real axis into a complex (imaginary) mass while the physical mode stays exactly protected, yet it does so through non-Hermiticity rather than a Ginsparg--Wilson relation. As a consequence its axial index vanishes through every available channel: the Hermitian trace $\Tr(\gfive D)$ (Karsten--Smit; Tiburzi), the bi-orthogonal index for non-Hermitian operators (Guo, Ma, and Zhang), and, closing the last loophole, the algebraic (Riesz-projector) index even when the operator is defective at an exceptional point; the last step extends Zenkin's theorem, that a chirally symmetric normal lattice Dirac operator has zero index, to non-normal and defective operators, which the \CDF operator requires. The \CDF operator therefore cannot reproduce the axial anomaly, not by a tunable near-miss but by an identity forced by the very chiral symmetry it preserves. We organize this and the neighboring constructions by the Dirac channel that carries the lattice structure: disorder in the $\g_\mu$ (chirality-preserving) channel realizes an Altland--Zirnbauer chiral class and can never carry an index, while disorder in the scalar channel, the coefficient of $\one$, breaks chiral symmetry and reproduces Callan--Harvey / domain-wall physics. A one-parameter family interpolates between the two, with the index a step function of the engaged channel: the 1983 operator and Kaplan's domain wall emerge as the two poles of a single disordered family, distinguished not by the disorder but by the channel that carries it. The picture unifies the lattice-QCD undoubling lineage with the tenfold-way classification of disordered Dirac fermions, and settles the fate of the disorder route to chiral lattice fermions.
\end{abstract}

\maketitle

\section{Introduction}
\label{sec:intro}

Species doubling is the central obstruction to putting chiral fermions on a lattice. Karsten and Smit tied it to the triangle anomaly~\cite{KarstenSmit}, and Nielsen and Ninomiya elevated it to a no-go theorem~\cite{NN1,NN2,NN3}: a lattice Dirac operator that is simultaneously local, translation invariant, Hermitian, and exactly chirally symmetric must double. The known escapes each relax one hypothesis---Wilson fermions give up chiral symmetry~\cite{Wilson}, Ginsparg--Wilson/overlap fermions realize it in a modified ($\anticomm{D}{\gfive}=aD\gfive R D$) form~\cite{GinspargWilson,Neuberger1,Neuberger2,Luscher}, and domain-wall fermions realize a single chiral mode as an edge state in an extra dimension~\cite{Kaplan,CallanHarvey}.

A less-travelled route relaxes \emph{Hermiticity} instead, keeping chiral symmetry exact. One of the earliest such proposals~\cite{Jacobs1983}---which we call \emph{chiral disordered fermions} (\CDF)---weighted the lattice derivative with a quenched, ultralocal random field, with the hope that the anomaly would re-emerge from the disorder average as a measure effect, in the spirit of Fujikawa~\cite{Fujikawa}. That program, and the cluster of non-Hermitian/disordered constructions around it~\cite{WeingartenVelikson,AlessandriniKrzywicki,AlonsoCortes,GrossRakow,Pelissetto}, was set aside in the late 1980s---refuted in instances but never \emph{explained}. Gross, Lepage, and Rakow found the Weingarten--Velikson model still doubled~\cite{GrossRakow}; Pelissetto found ghost states and a vanishing anomaly~\cite{Pelissetto}. The closest relative of the \CDF construction is the random-block lattice of Chiu~\cite{Chiu1988,Chiu1989,Chiu1990,Chiu1998proc}, in which the sites are the Cartesian product of random coordinates in each direction and the random spacings enter the lattice derivative as link weights. The randomness therefore sits in the $\g_\mu$ channel, exactly where the \CDF field sits, and the action is chirally and gauge invariant; unlike the \CDF operator it is anti-Hermitian, so that the determinant is positive and single-flavor dynamical simulations were possible. In two-dimensional QED on the random-block lattice, ensemble-averaged observables that do not depend on exact zero modes agree with the Schwinger-model solution on the torus~\cite{Chiu1998proc}, while in topologically nontrivial backgrounds the construction produces neither exact zero modes nor the correct index~\cite{ChiuZenkin1999}. In the language of Sec.~\ref{sec:channel} the random-block lattice is the normal member of the chiral class whose non-normal member is the \CDF operator; the two share the index obstruction and differ in the fate of their doublers (Sec.~\ref{sec:outlook}). The verdicts of the 1980s were verdicts on individual models, delivered before the Ginsparg--Wilson/overlap understanding existed. For the normal member of the class the failure was shown to be forced by Zenkin's theorem~\cite{Zenkin1998}, applied to the random-block lattice by Chiu and Zenkin~\cite{ChiuZenkin1999}; for the non-normal \CDF operator it required the bi-orthogonal machinery of non-Hermitian topology~\cite{MaZhang2024} and the defective extension of Sec.~\ref{sec:index}.

Here we supply the explanation. We prove that the disorder average of the \CDF operator~\cite{Jacobs1983} keeps $\anticomm{\Deff}{\gfive}=0$ exactly to all orders (Sec.~\ref{sec:exact}), that it lifts the doublers off the real axis without becoming Ginsparg--Wilson (Sec.~\ref{sec:spectrum}), and that its axial index vanishes in every sense---including, closing a loophole left open by the assumption of diagonalizability, at exceptional points (Sec.~\ref{sec:index}). We then show (Sec.~\ref{sec:channel}) that the decisive variable is the Dirac channel carrying the disorder, i.e.\ the Altland--Zirnbauer symmetry class~\cite{AltlandZirnbauer}, and that a single family interpolates between the \CDF operator and domain-wall fermions with the index as a step function of the channel. We stress at the outset that the general statement---exact chiral symmetry forbids the lattice anomaly---is not ours: Zenkin proved it for normal operators in 1998~\cite{Zenkin1998}, and Guo, Ma, and Zhang for the bi-orthogonal index of diagonalizable ones~\cite{GuoMaZhang}; our contribution is the disorder-averaged realization, the extension to non-normal and defective operators (which the \CDF operator requires, being non-normal), and the symmetry-class synthesis. Finally we ask whether any gauge-sector observable can stand in for the index and find that none can (Sec.~\ref{sec:gauge}): the topological susceptibility, the static quark potential, and the induced photon mass are all screened in the chiral channel as in the mass channel, because they couple to charge while the anomaly resides in chirality; they count charged modes, not their chirality. This leads us to two conjectures (Sec.~\ref{sec:outlook})---that the index is the \emph{unique} discriminator, and that the \CDF operator, made dynamical, would break the confining string by anomaly-free screening.

\section{The construction and its exact chiral structure}
\label{sec:exact}

We work in $d$ Euclidean dimensions with Hermitian $\g_\mu$, $\anticomm{\g_\mu}{\g_\nu}=2\delta_{\mu\nu}$, and $\gfive$. The \CDF operator of Ref.~\cite{Jacobs1983} is
\begin{equation}
D_\varphi=\sum_\mu \g_\mu M_\mu(\varphi)+m,\qquad
M_\mu(\varphi)=\delta^{s}_\mu+\tfrac12\,\hat\varphi\,(\delta^{+}_\mu-\delta^{-}_\mu),
\label{eq:Dphi}
\end{equation}
where $\delta^{s},\delta^{\pm}$ are the symmetric/forward/backward covariant differences, $\hat\varphi$ is the diagonal multiplication operator by a quenched Gaussian field with $\langle\varphi_n\rangle=0$, $\langle\varphi_n\varphi_m\rangle=\sigma\,\delta_{nm}$. The random field sits in the \emph{coefficient of $\g_\mu$}: every term of $D_\varphi$ at $m=0$ carries exactly one $\g_\mu$ and is a scalar in Dirac space otherwise. Hence, for \emph{every} realization,
\begin{equation}
\gfive\,D_\varphi\,\gfive=-D_\varphi \quad (m=0),
\label{eq:g5perreal}
\end{equation}
including in the presence of a gauge background, since the link variables enter $M_\mu$ and not the Dirac structure. Equation~\eqref{eq:g5perreal} gives $\gfive D_\varphi^{-1}\gfive=-D_\varphi^{-1}$ realization by realization, and averaging,
\begin{equation}
\anticomm{\Deff}{\gfive}=0\quad\text{exactly, to all orders in }\sigma,\;
\label{eq:exact}
\end{equation}
for the averaged effective operator $\Deff\equiv\langle D_\varphi^{-1}\rangle_\varphi^{-1}$. This is nonperturbative: it does not rely on a small-$\sigma$ expansion.

\section{Doubler lifting and the disorder spectrum}
\label{sec:spectrum}

The \CDF operator was designed to lift the doubler degeneracy, and it does; the analysis below makes the mechanism, and its limits, precise. Write $\Deff=D_0-\Sigma$, with $D_0$ the naive operator and $\Sigma$ the disorder self-energy generated by the vertex $V=\tfrac12\g_\mu\hat\varphi(\delta^{+}_\mu-\delta^{-}_\mu)$; to leading order $\Sigma=\langle V G_0 V\rangle$, $G_0=D_0^{-1}$. Two features follow.

\emph{The physical mode is untouched.} The vertex carries the second difference $\delta^{+}_\mu-\delta^{-}_\mu$, which vanishes at zero momentum. The protection is not a leading-order accident: every self-energy insertion, at any order in $\sigma$, terminates on an external vertex carrying that second difference at the external momentum, so $\Sigma(0)=0$ to all orders---the $p=0$ pole keeps its position and its bare mass for every $\sigma$.

\begin{figure}[htbp]
\centering
\includegraphics[width=\columnwidth]{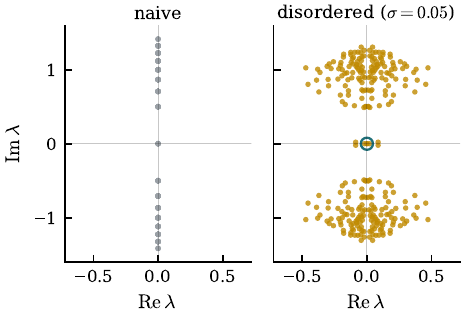}
\caption{Doubler lifting. \emph{Left}: the naive operator's spectrum lies on the imaginary axis, with $2^{d}\times2=8$ exact zero modes at the origin ($d=2$). \emph{Right}: the quenched disorder ($\sigma=0.05$) renders the operator non-Hermitian and pushes the eigenvalues---in particular the doubler zero modes---into the complex plane (an imaginary mass), while the two physical $p=0$ modes remain exactly protected at the origin (circled) and the $\lambda\to-\lambda$ symmetry of Eq.~\eqref{eq:g5perreal} is preserved.}
\label{fig:spectrum}
\end{figure}

\emph{The doublers are pushed off the real axis.} The exact relation $\anticomm{\Deff}{\gfive}=0$ forbids a scalar (mass) self-energy in the chiral limit---the doublers cannot be given a real Wilson mass. What the disorder generates instead is a \emph{purely imaginary} vector self-energy: writing $\Deff(p)=i\g_\mu K_\mu(p)$, the $K_\mu$ acquire imaginary parts at the doubler momenta, and the doubler poles move into the complex plane---the ``large imaginary mass'' anticipated in Ref.~\cite{Jacobs1983}; we call these the imaginary-mass fermions (\IMF) of the doubler sector. The effect is sharp: in $d=2$ the naive operator has $2^{d}\times2=8$ exact zero modes (the $2^{d}=4$ momenta $\{0,\pi\}^{2}$, each doubly degenerate in Dirac space), and the disorder lifts all but the two physical $p=0$ modes (Fig.~\ref{fig:spectrum}). The naive spectrum lies on the imaginary axis; the disorder pushes it into the complex plane while preserving the exact $\lambda\to-\lambda$ symmetry that is the fingerprint of Eq.~\eqref{eq:g5perreal}. At $m\neq0$ there is in addition a Wilson-like scalar self-energy $\propto\sigma m\,\hat p^{2}$ that splits the doublers from $p=0$; being proportional to $m$, it is irrelevant in the chiral limit.

This is genuine undoubling---the physical mode isolated at the origin, the doublers gapped off the real axis---but it is not of Ginsparg--Wilson type. A GW operator obeys $\anticomm{D}{\gfive}=aD\gfive RD\neq0$~\cite{GinspargWilson}; here $\anticomm{\Deff}{\gfive}=0$. The \CDF operator evades Nielsen--Ninomiya through non-Hermiticity, not through the soft chiral breaking of the overlap---and that distinction is exactly what decides the anomaly (Sec.~\ref{sec:index}): the doublers are lifted, but by a chirality-preserving mechanism, and a chirality-preserving lifting carries no index.

\section{The index and the anomaly}
\label{sec:index}

We ask Wilson's question\footnote{On being shown an early draft of~\cite{Jacobs1983}, K.~G.~Wilson remarked that the first thing he would do would be to check that the approach recovered the anomaly; see the Acknowledgments.}: does the \CDF operator reproduce the axial anomaly? In a background of topological charge $Q$ the continuum index theorem sets the net chirality of the zero modes equal to $Q$. On the lattice the answer is negative through every channel.

\emph{Hermitian.} From Eq.~\eqref{eq:exact}, $\Tr(\gfive \Deff)=-\Tr(\gfive\Deff)=0$; the doubler species carry compensating anomalies, the lattice restatement of Karsten--Smit~\cite{KarstenSmit} made explicit for exactly-chiral actions by Tiburzi~\cite{Tiburzi}.

\emph{Bi-orthogonal.} For a non-Hermitian operator the relevant object is the bi-orthogonal index $Q_L=n_--n_+$ built from left/right eigenvectors~\cite{Biorthogonal}. Guo, Ma, and Zhang proved that $\anticomm{D}{\gfive}=0$ forces $Q_L=0$~\cite{GuoMaZhang}: each nonzero mode has vanishing bi-orthogonal chirality $\langle L_i|\gfive|R_i\rangle$, and $\sum_i\langle L_i|\gfive|R_i\rangle=\Tr\gfive=0$. We confirm this for the disorder-averaged operator~\eqref{eq:Dphi} in $Q=1,2$ backgrounds.

\emph{Exceptional points.} The bi-orthogonal argument assumes a complete eigenbasis and breaks down at exceptional points, where $\langle L_i|R_i\rangle\to0$ and $D$ is defective. The natural index there is the algebraic one,
\begin{equation}
\mathrm{ind}\,D=\Tr(\gfive P_0),\qquad P_0=\frac{1}{2\pi i}\oint_{0}(z-D)^{-1}\,dz,
\end{equation}
which is defined for defective $D$. We close the loophole:

\begin{theorem}
Let $D$ satisfy $\anticomm{D}{\gfive}=0$ on a finite lattice---so that $D$ acts on a finite-dimensional space and $\Tr\gfive=0$---with $\gfive^2=1$. Then $\Tr(\gfive P_\lambda)=0$ for every $\lambda\neq0$, and hence $\Tr(\gfive P_0)=0$, for any gauge background---whether or not $D$ is diagonalizable.
\end{theorem}
\begin{proof}
The Riesz projectors satisfy $\gfive P_\lambda\gfive=P_{-\lambda}$ (from $\gfive D\gfive=-D$) and $P_\lambda P_{-\lambda}=0$ for $\lambda\neq0$. Then $\Tr(\gfive P_\lambda)=\Tr(P_\lambda\gfive P_\lambda)=\Tr(P_\lambda P_{-\lambda}\gfive)=0$. Since $\sum_\lambda P_\lambda=\one$ and $\Tr\gfive=0$, $\Tr(\gfive P_0)=0$.\qquad
\end{proof}

The tempting escape---an odd-length Jordan chain at $\lambda=0$, whose chirality alternates and appears to net $\pm1$---is self-defeating: every nonzero eigenspace is chirality-neutral, so any imbalance at $\lambda=0$ is compensated within the $\lambda=0$ generalized eigenspace. Numerically, explicit defective chiral operators (Jordan blocks, eigenvector condition number up to $10^{292}$) and $\sim\!10^3$ random near-defective ones all give $|\mathrm{ind}\,D|\lesssim10^{-14}$. This strengthens the no-go of Ref.~\cite{GuoMaZhang} to the non-diagonalizable case: non-Hermitian spectral degeneracy cannot be engineered to smuggle in a chiral index. The closure is specific to the chiral (axial) index; a defective operator may still carry genuine point-gap non-Hermitian topology, but that is not the anomaly. The vanishing is a property of the chiral \emph{class}, not of non-Hermiticity as such: in the complementary pseudo-Hermitian, real-spectrum regime, non-Hermitian Dirac operators do carry nonzero indices~\cite{roy2025zero}; together, the two regimes bracket the non-Hermitian question.

The theorem has two antecedents. Zenkin~\cite{Zenkin1998} proved in 1998, as an extension of the Nielsen--Ninomiya theorem, that any chirally symmetric \emph{normal} Dirac operator has zero index, on regular or random lattices and whether or not the operator is translation invariant or local; the only hypothesis is normality. Guo, Ma, and Zhang~\cite{GuoMaZhang} proved the statement for the bi-orthogonal index of a diagonalizable operator. The theorem above removes both hypotheses. The extension is not idle here: the \CDF operator is not normal for any $\sigma>0$ (we find $\|[D_\varphi,D_\varphi^\dagger]\|/\|D_\varphi\|^2=0.02$ to $0.06$ for $\sigma=0.01$ to $0.2$ on a $16\times16$ lattice, and $0.23$ for $\Deff$ at $\sigma=0.05$; the naive operator at $\sigma=0$ is anti-Hermitian and hence normal), so Zenkin's theorem does not apply to it and the case is closed only by the non-normal extension.

The trace identity $\Tr\gfive=0$ that drives the proof has a Ginsparg--Wilson counterpart, the chirality sum rule of Chiu~\cite{Chiu1998zero}: the total chirality of all eigenmodes of a Ginsparg--Wilson operator vanishes, and the chirality of its zero modes is balanced by real modes at $\lambda=2/a$ of definite chirality. The two statements are the same identity with the compensating chirality in different places. For an exactly anticommuting operator it must lie inside the $\lambda=0$ generalized eigenspace, which forces the index to vanish; for a Ginsparg--Wilson operator it lies at the cutoff, which leaves the physical zero modes free to carry it. Chiu and Zenkin~\cite{ChiuZenkin1999} put the same point in terms of the chirally symmetric kernel: writing a Ginsparg--Wilson operator as $D=D_c(1+aRD_c)^{-1}$, a nonzero index requires $\det(1-aRD)=0$, that is, $D_c$ must be singular in nontrivial sectors, while the subtracted propagator $(D_c+m)^{-1}$ remains finite~\cite{Chiu1999propagator}; the index and the anomaly are moreover independent of $R$ and fixed by $D_c$ alone~\cite{Chiu1999index}. A chirally symmetric $D_c$ that exists on every configuration, which is what the \CDF operator is, has zero index: by Zenkin's theorem when $D_c$ is normal, and by the Theorem above in general, including the \CDF operator.

It is worth marking where finite-dimensionality enters. The final step of the proof is $\Tr\gfive=0$, which is precisely the finite-lattice statement; the continuum evades it because Fujikawa's Jacobian is a \emph{regularized} $\Tr\gfive$~\cite{Fujikawa}---which is where the anomaly lives. This localizes why the original hope of Ref.~\cite{Jacobs1983}, that the anomaly would re-emerge from the disorder average as a measure effect, could not be realized: on a finite lattice with $\anticomm{D}{\gfive}=0$ exact, the measure has nowhere to put a Jacobian.

\section{The symmetry-class criterion and its two limits}
\label{sec:channel}

\subsection{The channel is the control parameter}

The results above are one face of a single organizing statement: what decides anomaly reproduction is the Dirac channel that carries the lattice structure---equivalently, its Altland--Zirnbauer symmetry class~\cite{AltlandZirnbauer}. Write a general structured operator as $D=\g_\mu K_\mu(x)+\Phi(x)\,\one$. Structure in the $\g_\mu$ (vector) channel preserves $\anticomm{D}{\gfive}=0$---a chiral (AIII/BDI-type) class---and by the Theorem carries no index in any sense. Structure in the scalar channel, the coefficient of $\one$, is the mass channel: it breaks chiral symmetry and is thereby \emph{permitted} an index; whether it realizes one depends on the profile of $\Phi$. Permitting an index is not the same as carrying one. Chiu~\cite{Chiu1999construction,Chiu2001index} constructed a Ginsparg--Wilson operator, $\gfive$-Hermitian, exponentially local, and free of doublers, that reproduces the topological charge density as its axial anomaly in weak-coupling perturbation theory~\cite{ChiuHsieh2002} and yet has no topological zero modes in nontrivial backgrounds satisfying a mild condition [Eq.~(30) of Ref.~\cite{Chiu2001index}]. Within the mass channel, a correct local anomaly and exact zero modes are therefore distinct properties; one expects such an operator to reproduce trivial-sector observables and to fail where zero modes enter, in the 't~Hooft vertex and the $\eta'$ mass. The channel criterion locates the \CDF operator on the side where the index is forbidden; on the permitted side, the overlap operator and domain-wall fermions at $N_s\to\infty$ are the constructions known to have the zero modes, and Chiu's example shows that the class also contains members that do not.

A one-parameter family with the disorder held fixed in the $\g_\mu$ channel and the doubler-gapping $\gfive$-Hermitian channel dialed on by $\lambda$ makes the index a step function of $\lambda$ (Fig.~\ref{fig:latch}): $0$ in the chiral class of the \CDF operator, and $Q$ once the anomaly-carrying channel engages. The \emph{same} disorder gives opposite answers in the two channels; it is the channel, not the disorder, that decides.

\subsection{Domain-wall fermions as the mass-channel limit}
\label{sec:wall}

\begin{figure}[t]
\centering
\includegraphics[width=\columnwidth]{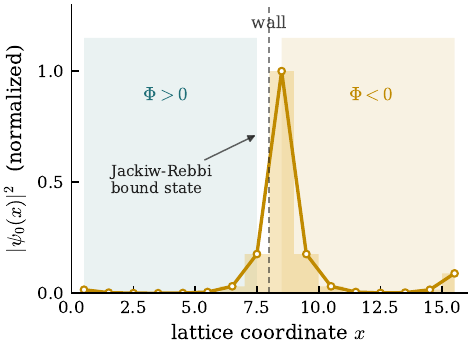}
\caption{The mass-channel limit is domain-wall physics. With the doublers gapped by a Wilson term, a sign-changing mass $\Phi(x)$ (kink--antikink) binds a near-zero mode whose density $|\psi_0(x)|^2$ localizes at the walls (Callan--Harvey/Jackiw--Rebbi). The disordered mass channel generalizes this to a wall at every zero crossing---random-mass Dirac, the ``disorder-induced'' domain wall.}
\label{fig:wall}
\end{figure}

The mass channel is not an abstract foil: its natural realizations are Wilson, domain-wall, and overlap fermions. A position-dependent mass $\Phi(x)$ that changes sign is a domain wall, and at the wall the Callan--Harvey mechanism~\cite{CallanHarvey} binds a chiral zero mode---the content of Kaplan's construction~\cite{Kaplan}, whose reduction along the extra dimension is the overlap operator~\cite{Neuberger1,Luscher}. We verify the binding directly: with the doublers gapped by a Wilson term, a sign-changing $\Phi$ localizes a near-zero mode at the wall (Fig.~\ref{fig:wall}), and the chirality-weighted count of such modes returns $Q$ in a charge-$Q$ background. The \emph{disordered} mass channel is then nothing but the random-mass Dirac problem: a mean-zero random $\Phi$ seeds a domain wall wherever it crosses zero, each binding a Jackiw--Rebbi mode, with the Dyson density-of-states singularity as the collective fingerprint~\cite{Dyson,RandomMass}. In this precise sense the disorder framework \emph{contains} domain-wall fermions---as ``disorder-induced walls''---in its $\gfive$-Hermitian channel, exactly as it contains the exactly-chiral, anomaly-free \CDF operator in its non-Hermitian channel.

\begin{figure}[t]
\centering
\includegraphics[width=\columnwidth]{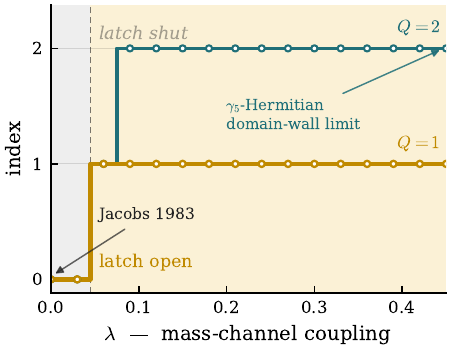}
\caption{The anomaly latch. With the disorder fixed in the $\g_\mu$ channel, the bi-orthogonal index is a step function of the coupling $\lambda$ to the $\gfive$-Hermitian (mass) channel: $0$ in the chiral (Altland--Zirnbauer) class of the \CDF operator, and $Q$ once the anomaly-carrying channel engages. Data for $Q=1,2$ on a $16\times16$ lattice.}
\label{fig:latch}
\end{figure}

The comparison sharpens the criterion. Both constructions lift the doublers: the mass channel by a real Wilson/wall gap, the \CDF operator by the purely imaginary vector self-energy of Sec.~\ref{sec:exact}. But doubler-lifting is \emph{necessary, not sufficient} for the anomaly---the lifting must occur in the chiral-symmetry-breaking channel. The mass wall qualifies; the vector disorder does not. The latch of Fig.~\ref{fig:latch} is precisely the crossover between the two, with the $\lambda=0$ end the \CDF operator and the far end domain-wall fermions: one family, one knob, two symmetry classes.

\section{Gauge-sector observables and the shadow of the anomaly}
\label{sec:gauge}

The index of Sec.~\ref{sec:index} is a chirality-resolved quantity. It is natural to ask whether some more accessible, gauge-invariant observable---one that does not require the eigenmodes' chirality---could serve as a proxy and expose the missing anomaly. It cannot: the standard probes of the gauge sector are screened in the chiral channel as in the mass channel, because they couple to electric charge, and the \CDF doublers are charged; what they measure is how much long-range charge there is, not its chirality. We illustrate in the two-dimensional Schwinger model~\cite{Schwinger}, sampled by Monte Carlo.

\emph{Topological susceptibility.} With dynamical massless fermions the anomaly screens topological charge, $\chi_t\to0$. So does the chiral-channel operator: the fermion determinant is suppressed on $Q\neq0$ backgrounds whenever the Dirac operator has near-zero modes, and both channels have $\sim|Q|$ of them (the mass channel with net chirality $Q$, the chiral channel in canceling $\pm$ pairs). $\chi_t$ counts near-zero modes, not their chirality, and vanishes in both.

\begin{figure}[t]
\centering
\includegraphics[width=\columnwidth]{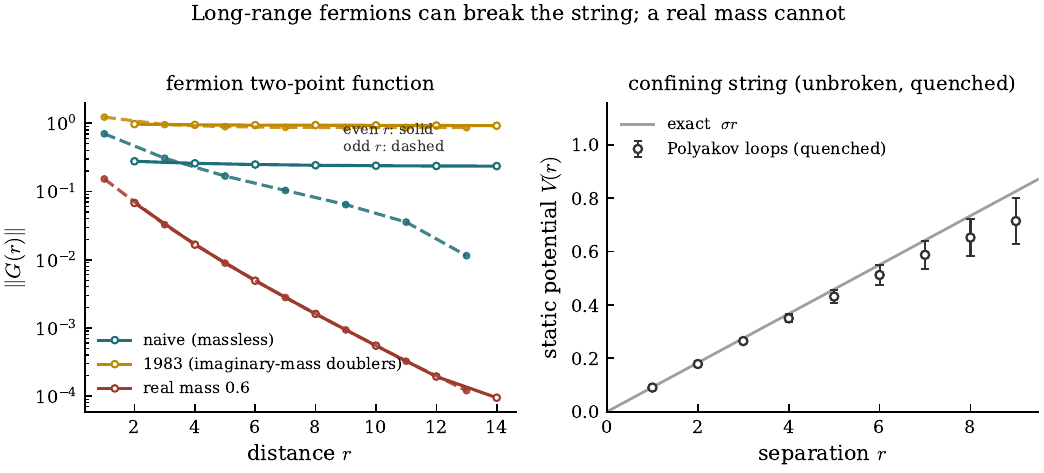}
\caption{Why the gauge sector is blind. \emph{Left}: fermion two-point function $\|G(r)\|$ (even $r$ solid, odd $r$ dashed). A real mass (red) decays exponentially and decouples; the naive massless mode and the \CDF (\IMF) doublers both stay long-range and can screen. \emph{Right}: the quenched Polyakov static potential---the unbroken confining string, matching the exact $\sigma r$. Long-range charged fermions break this string in either channel, independently of the anomaly.}
\label{fig:string}
\end{figure}

\emph{The static potential.} Pure two-dimensional compact U(1) confines at every coupling, with a linear static potential $V(r)=\sigma r$ and exactly known string tension $\sigma=-\log[I_1(\beta)/I_0(\beta)]$; we recover it from Polyakov-loop correlators (Fig.~\ref{fig:string}, right). With dynamical massless fermions the charge is screened and the string breaks; the two-point functions of Fig.~\ref{fig:string} (left) show that the \CDF (\IMF) doublers, like the naive massless mode, remain long-range and can therefore screen in exactly the same way. String breaking in the dynamical \CDF theory is thus the expected outcome in both channels---though, unlike the quenched tension, we have not simulated it; it is part of Conjecture~2 below.

\emph{The induced photon mass.} The Schwinger mass $m_\gamma^2=e^2/\pi$ is the sharpest gauge-sector fingerprint of the anomaly. Measuring the transverse vacuum polarization $\Pi_T(0)$ as the curvature of the fermion effective action, the overlap operator reproduces the one-flavor value, the naive operator gives a doubler-inflated multiple, and the \CDF operator gives---from robust estimators over the disorder ensemble---a positive value comparable to the naive one. It, too, screens. The mechanism is visible in the fermion two-point function (Fig.~\ref{fig:string}, left): a genuinely massive mode decays exponentially and decouples, whereas the \CDF (\IMF) doublers remain \emph{long-range} and feed the long-distance vacuum polarization; their $(-1)^r$ oscillation trims $\Pi_T(0)$ but does not cancel it. (We caution that the determinant-curvature estimator is heavy-tailed for the non-Hermitian operator---its mean is dragged toward zero by rare near-zero-mode configurations---so that a definitive value of $\Pi_T(0)$ awaits the dedicated current--correlator computation of Sec.~\ref{sec:outlook}; the robust center, however, is clearly nonzero.)

Three independent gauge-sector probes---topological susceptibility, the static potential, and the photon mass---thus all fail to separate the anomaly-free chiral construction from a genuinely anomalous one. The reason is uniform: every gauge/vector observable couples to charge; the doublers carry charge and, here, stay long-range; and the axial anomaly resides not in the presence of these modes but in their chirality. The index resolves chirality; the gauge sector does not. The gauge sector can tell operators apart by how many long-range charged modes they carry (it separates the \CDF and naive operators from the overlap operator by the value of $\Pi_T(0)$), but not by the chirality of those modes; the chirality-resolved index of Sec.~\ref{sec:index} is, in this sense, the only probe of the index itself---a point we sharpen into a conjecture below.

\section{Discussion}
\label{sec:discussion}

It is worth stating the relation to prior work plainly. That exact chiral symmetry at finite lattice spacing precludes the axial anomaly is, in the Hermitian setting, long understood: implicit in the doubler cancellation of Karsten and Smit~\cite{KarstenSmit} and made explicit for minimally-doubled actions by Tiburzi~\cite{Tiburzi}. That any chirally symmetric normal operator, local or not, on a regular or random lattice, has zero index is Zenkin's extension of the Nielsen--Ninomiya theorem~\cite{Zenkin1998}; its form for the bi-orthogonal index of a diagonalizable non-Hermitian operator is due to Guo, Ma, and Zhang~\cite{GuoMaZhang}, whose general argument and two-dimensional check we reproduce. Our own additions are deliberately narrow: the disorder-averaged \CDF operator~\cite{Jacobs1983} satisfies $\anticomm{\Deff}{\gfive}=0$ exactly and nonperturbatively; the no-go survives the loss of diagonalizability, so that non-Hermitian degeneracy cannot evade it (Sec.~\ref{sec:index}); and the whole family---chiral and mass channels alike---is organized by its Altland--Zirnbauer symmetry class.

The physical conclusion is clarifying rather than merely negative. The disorder route \emph{does} lift the doublers---pushing them off the real axis into a complex mass---but it cannot regulate the anomaly, and the reason is the exact chiral symmetry it was built to preserve. Undoubling and anomaly reproduction are distinct requirements: the first is met by any lifting mechanism, the second only by one that breaks chiral symmetry in the $\gfive$-Hermitian channel. A chirality-preserving lifting, however ingenious, carries no index.

Read this way the result is a unification. The lattice-QCD undoubling lineage---Karsten--Smit~\cite{KarstenSmit}, Nielsen--Ninomiya~\cite{NN1}, Wilson~\cite{Wilson}, Ginsparg--Wilson and overlap~\cite{GinspargWilson,Neuberger1,Luscher}, Kaplan's domain wall~\cite{Kaplan} and Callan--Harvey inflow~\cite{CallanHarvey}---and the condensed-matter classification of disordered Dirac fermions---the Altland--Zirnbauer classes~\cite{AltlandZirnbauer}, random-mass Dirac with its Jackiw--Rebbi modes and Dyson singularity~\cite{Dyson,RandomMass}---are two views of one criterion: which Dirac channel carries the structure fixes the symmetry class, and the class fixes whether an index, and hence the anomaly, can appear.

There is also a circle closing here. The construction of Ref.~\cite{Jacobs1983} and Kaplan's domain wall~\cite{Kaplan} enter this family as its two poles: the same quenched random field, placed in the $\g_\mu$ channel, gives the exactly chiral, anomaly-free \CDF operator; placed in the scalar (mass) channel, it seeds domain walls at its zero crossings and gives the anomaly-carrying construction that succeeded. The two are not rival tricks but the two exact limits of one disordered operator---the two ways of splitting the Nielsen--Ninomiya hypotheses, non-Hermiticity against soft chiral breaking---distinguished by nothing except the Dirac channel that carries the field. What was proposed in 1983 as a route to chiral lattice fermions returns, four decades on, as the chiral-class pole of the family whose mass-channel pole is the route that worked.

\section{Conjectures and outlook}
\label{sec:outlook}

Two statements reach beyond what we have proved. We set them out as conjectures, each with its supporting evidence and the calculation that would settle it.

\emph{Conjecture 1 (gauge-sector blindness).} No gauge-invariant, chirality-blind observable distinguishes two lattice fermions with the same long-range charged spectral content, whatever their indices; in particular, none detects the vanishing of the index of a chiral-class operator. The statement must be read with care. The photon mass does separate the \CDF operator from the overlap operator by its value, a result close to the naive one against $e^2/\pi$ (Sec.~\ref{sec:gauge}), but it separates the naive operator in the same way and for the same reason: $\Pi_T(0)$ counts the long-range charged modes with the same sign for every species, whereas the anomaly counts them with their chirality, and the \CDF doublers are long-range. The three probes of Sec.~\ref{sec:gauge} are evidence for the conjecture in this form, and the heuristic is sharp: any such observable factors through the charged spectral density, which the index leaves invariant. A proof would make that factorization precise. The cleanest test is the Ginsparg--Wilson operator of Ref.~\cite{Chiu2001index}, which shares the local anomaly of the overlap operator and differs from it only in the absence of topological zero modes: the conjecture predicts the same $\Pi_T(0)$ for the two, with the index as the only quantity that tells them apart.

\emph{Conjecture 2 (anomaly-free deconfinement).} Made dynamical, the \CDF operator breaks the confining string of the two-dimensional gauge theory---it screens the static charge and gives the photon a mass---\emph{without} an axial anomaly. This would be a confining theory deconfined by anomaly-free charged matter, with no counterpart among Wilson, overlap, or real-mass fermions, for which confinement-breaking and the anomaly are welded together. The naive operator is not a counterexample: it, too, deconfines with zero \emph{net} anomaly (Fig.~\ref{fig:string}), but only by being four massless anomalous fermions in disguise, their anomalies canceling pairwise~\cite{KarstenSmit}. The \CDF theory is undoubled---a single protected massless pole---and yet the screening persists, carried by modes that are neither massless nor decoupled: screening without either masslessness or an anomaly. That is its distinguishing ingredient, the imaginary-mass doubler spectrum: neither massless nor conventionally massive, long-range enough to screen (Fig.~\ref{fig:string}) yet chirally trivial. The long-range two-point function and the robust positivity of $\Pi_T(0)$ support it. The random-block lattice supplies the control. There the doubled modes acquire masses of order $1/a$ because their chiral symmetry is spontaneously broken, and the associated Goldstone boson decouples with them~\cite{Chiu1988}; this is not in conflict with Sec.~\ref{sec:spectrum}, where $\anticomm{\Deff}{\gfive}=0$ forbids a scalar self-energy in the averaged operator, and it names the real difference between the two members of the class: spontaneous breaking in the normal ensemble against an imaginary mass in the non-normal one. The ensemble-averaged propagator in the free-field limit and the vacuum polarization tensor in weak-coupling perturbation theory agree with their continuum values beyond about three average lattice spacings~\cite{Chiu1988,Chiu1989}; and with dynamical fermions on $64$ random-block lattices of size $12\times12$, the chiral correlation function agrees with the exact Schwinger-model solution on the torus~\cite{Chiu1998proc}. The two constructions share the symmetry class and the index obstruction but not the fate of their doublers, so Conjecture~2 is a statement about the \CDF realization and its long-range \IMF sector, not about the chiral class as a whole. Since both lattices also carry a protected massless physical mode, which in two dimensions screens by itself, a side-by-side comparison must isolate the doubler contribution, for instance the $(-1)^r$ component of Fig.~\ref{fig:string}.

Two caveats frame Conjecture~2, and both are old acquaintances. The first is reflection positivity. The imaginary-mass doublers are precisely the modes Pelissetto identified as ghosts~\cite{Pelissetto}, and Alonso and Cort\'es already observed that the simplest chirally invariant undoubled actions violate Osterwalder--Schrader positivity at the ultralocal level~\cite{AlonsoCortes}. A dynamical theory whose screening is carried by complex-mass poles must therefore either restore positivity in its physical sector or accept that the ``deconfinement'' diagnoses the non-standard reality of the \IMF sector rather than a new physical phase; the conjecture asserts the behavior of the observables, and we deliberately leave its interpretation open. The second is the measure. With U(1) links, $\det D_\varphi$ is complex in general---the $\lambda\to-\lambda$ pairing does not by itself enforce reality---so the dynamical theory carries a phase problem, on top of the quenched-versus-annealed choice for the disorder average of the determinant. (Ma and Zhang obtain a positive partition function for \emph{paired} flavors of their one-sided non-Hermitian fermions~\cite{MaZhang2024}; whether an analogous pairing tames the \CDF measure is open.) It is a virtue of the calculation proposed below that it requires neither: the disorder-connected correlator is evaluated on quenched ensembles and sidesteps the dynamical measure entirely.

Both conjectures turn on a single object: the disorder-averaged, disorder-connected current--current correlator $\langle j_\mu(x)\,j_\nu(0)\rangle$, evaluated with the conserved lattice current and with an analytic handle on the imaginary-mass fermion loop, extrapolated in momentum and quark mass. Its zero-momentum transverse part is $\Pi_T(0)$; whether the $(-1)^r$-oscillating doubler contribution survives that sum decides at once whether the string breaks (Conjecture 2) and whether any gauge observable can (Conjecture 1). We regard this as the natural next calculation---and as the point at which the imaginary-mass structure of the \CDF operator does something that neither a naive nor a Ginsparg--Wilson operator can. It should be computed side by side on the \CDF and random-block lattices: identical screening would place the effect in the chiral class, different screening in the imaginary-mass structure.

\vspace{.3cm}

\begin{acknowledgments}
I am grateful to Ken~Wilson, who, on being shown the construction of Ref.~\cite{Jacobs1983} remarked that the first thing he would check was whether it recovered the anomaly---a question that I, four decades later, now understand, and answer in the negative. I thank Ting-Wai Chiu for pointing out the random-block lattice, Zenkin's theorem, and the chirality sum rule, and for copies of the papers concerned.
\end{acknowledgments}

\emph{Data availability.}---All numerical results are reproduced by the accompanying computational backbone~\cite{Zenodo}.


\bibliographystyle{apsrev4-2}
\bibliography{undoubling-2}

\end{document}